\documentclass[preprint,prl,aps,showpacs]{revtex4}

\usepackage{bm,here}
\usepackage{graphicx}
\usepackage{amsmath,amsxtra,amsfonts,amssymb,amstext,latexsym}
\usepackage{times}
\usepackage{mathptm}

\input{definitions.def}
\begin{document}


\title{Impurity and Trace Tritium Transport in Tokamak Edge Turbulence}
\author{V.~Naulin}
\affiliation{Association EURATOM-Ris{\o} National Laboratory,
OPL-128 Ris{\o}, DK-4000 Roskilde, Denmark}
\date{\today}

\begin{abstract}
The turbulent transport of  impurity or minority species, as for example Tritium, is
investigated in  drift-Alfv\'en
edge  turbulence. The full effects of perpendicular and parallel convection
are kept for the impurity species.
The impurity density  develops a granular structure with steep
gradients and locally exceeds  its initial values due to the compressibility of the flow. 
An approximate decomposition of the impurity flux into a diffusive
part and an effective convective part (characterized by a pinch velocity)
is performed and a net inward pinch effect is recovered. The pinch
velocity  is explained in terms of Turbulent Equipartition
\cite{Nycander:Yankov:1995} and  is found to vary poloidally.
The results show that impurity transport modeling needs to be
two-dimensional, considering besides the radial direction  also the
strong poloidal variation in the
transport coefficients.
\end{abstract}

\pacs{
      52.25.Gj, 
      52.35.Ra, 
      52.65.Kj  
}
\maketitle

The transport properties of impurities are of great concern in magnetic
fusion devices. In particular impurities
accumulate  in the center of the plasma, where they are responsible
for significant radiation losses. It is well established that turbulence is the key drive for
plasma  transport in the edge region and thus turbulence will also
dominate the transport of impurities in that region.   
 While in neutral fluids the transport of passive
scalars is a very active field of research
\cite{Falkovich:Gawdzki:Vergassola:2001}, it has in the context of
plasma turbulence not yet found corresponding attention and
measurements of impurity transport are most often
interpreted within reduced 1D transport models \cite{Dux:2003,Puiatti:etal:2003}.
In plasma turbulence the velocity field is in general compressible
what makes the transport and mixing more complex than in
incompressible flows \cite{Falkovich:Gawdzki:Vergassola:2001}. This
puts strong demands on the numerical modeling.

Pinching of impurities has been observed as well as fast inwards
transport of trace Tritium in JET Tritium puffs \cite{Zastrow:1999}.
Especially in the edge region an inward convective flux (pinch) of
impurites is found, which is proportional to the measured diffusion
\cite{Dux:2003}.  In the core where anomalous, turbulent transport is less
important than in the edge  the observed inward pinch of
impurities seems to be in good agreement with neoclassical
predictions based on trapped particle orbits. 
The Pfirsch-Schl{\"u}ter impurity
transport in the edge \cite{Rutherford:1974}, leads to an inward
impurity pinch due to ion-impurity collisions.
However, there  is  no  
explanation for the high inward pinch velocities needed to describe impurity transport
through in the  turbulent edge region and it's scaling with the
effective diffusion. 

Here we investigate the transport of impurities as a passive tracer
field in electromagnetic edge turbulence described by a standard model
of   drift-Alfv{\'e}n turbulence
~\cite{Scott:1997:1,Scott:1997:2,Naulin:2003}. 
Flux tube geometry is used,  with
local slab-like coordinates $\left(x,y,s\right)$~\cite{Scott:2001}.
The following equations for the fluctuations in density $n$, potential
$\phi$ with associated vorticity $\omega = \nabla_\perp^2 \phi$, current
$J$ and parallel ion velocity $u$ arise in the usual drift-scaling:
\begin{subequations}
\begin{gather}
\frac{\partial\omega}{\partial t} + \{ \phi, \omega\} =
\mathcal{K}\left( n \right) + \nabla_\shortparallel J 
+ \mu_\omega\nabla_{\perp}^{2}\omega, \label{eq:eqvor} \\
\frac{\partial n}{\partial t} + \{ \phi, n_{EQ} + n \} = 
\mathcal{K}\left( n  - \phi \right) +  \nabla_\shortparallel
\left( J  -  u \right) + \mu_{n}\nabla_{\perp}^{2} n , \label{eq:eqne} \\
\frac{\partial}{\partial t} \left( \widehat{\beta}A_{\shortparallel} +
\widehat{\mu} J \right) + \widehat{\mu}\left\{ \phi , J \right\} = 
\nabla_\shortparallel \left( n_{EQ} + n - \phi \right) - C J , \label{eq:eqpsi} \\
\widehat{\eps}\left(\frac{\partial u}{\partial t} +
\left\{ \phi , u \right\} \right) = - \nabla_\shortparallel
\left( n_{EQ} + n \right) . \label{eq:equi} \end{gather}
\end{subequations}
In addition to the cross-field advection, the parallel derivatives
carry non-linearities entering through $A_{\shortparallel}$, while the
operator $\mathcal{K}$ represents effects of normal and geodesic
curvature due to magnetic field inhomogeneity with $\omega_B = 2
\frac{L_\perp}{R}$, R being the tokamak major radius and $L_\perp$ 
the mean gradient length of the pressure profile,
\[
\nabla_{\shortparallel} n = \frac{\partial n}{\partial s} -
  \widehat{\beta}\left\{ A_{\shortparallel},n \right\} ,   \qquad
\mathcal{K}(n) = - \omega_B\left( \sin s\,\frac{\partial n}{\partial x} +
                   \cos s\,\frac{\partial n}{\partial y} \right) .
\]
The parallel current $J$ is connected to the magnetic potential given by
$J=-\nabla_{\perp}^{2}A_{\shortparallel}$.
The parameters reflect the competition between
parallel and perpendicular dynamics, represented in the scale ratio
$\widehat{\eps}=(qR/L_\perp)^2$ where $q$ is the safety factor. The electron parallel
dynamics is controlled by the parameters
\begin{equation} \label{eq:eqparms}
  \widehat{\beta} = \frac{2\mu_0 p_{e,0}}{B^2}\,\widehat{\eps} , \qquad
  \widehat{\mu} = \frac{m_e}{M_i}\,\widehat{\eps} , \qquad
  C = 0.51\,\frac{L_\perp}{\tau_e c_s}\,\widehat{\mu} = \nu\widehat{\mu} ,
\end{equation}
with $\tau_e$ being the electron collision time and the factor $0.51$
reflects the parallel resistivity~\cite{Braginskii:1965}. $n_{EQ}$ is an
equilibrium density associated with corresponding  neoclassical fields and currents.
In detail the  curvature operator on the equilibrium
density balances  the neoclassical Pfirsch-Schl\"uter current
$J_{PS}$
\begin{equation}
\label{eq:equil1}
\mathcal{K}\left( n_{EQ} \right) +
\nabla_\shortparallel J_{PS} = 0\;,
\end{equation}
 where the Pfirsch-Schl\"uter current is
driven by the corresponding neoclassical potential
\begin{equation}
\label{eq:equil2}
\nabla_\shortparallel
\left(\phi_{PS} \right)  =  C J_{PS} \,.
\end{equation}

To determine the evolution of the impurity density $n_{imp}$ 
we  assume that the contribution of the impurities to the gross plasma
density $n_{i,0}$ is negligible, i.~e.~$n_i = n_{i,0} + n_{imp} \approx
n_{i,0}$. 
In experiments the  assumption $n_{i,0} \gg n_{imp}$ 
is possibly not always fulfilled, especially not  during  the
initial phase of a Tritium gas puff, where locally in the edge region
the Tritium migth contribute a significant part of overall plasma
density.
The bulk plasma is quasineutral  with $n_e =
n_{i,0}$, allowing to regard either the ion- or
electron continuity equation to determine the density evolution.  
We can, however, not use a corresponding relation for $ n_{imp}$.
For cold impurities the drift velocity is given by
the $E \times B$-  and the ion polarisation drift:
\begin{equation}
d_{t} n_{imp} 
=  \frac{M}{Z\widehat{\eps}} \nabla_{\perp}\cdot \left( n_{imp} d_{t} \nabla_{\perp} \phi  \right) 
- n_{imp} \mathcal{K}\left( \phi \right) 
- \nabla_{\|} \left( n_{imp} u \right) - \mu_{imp} \nabla_{\perp}^{2} n_{imp}
\label{eq:impurity}
\end{equation}
Here  we have introduced the relative mass of
the impurities $M = M_{imp}/M_{i}$ and $Z$ indicates the charge
state of the impurity ions. 
The fluctuating quantities $\phi$ and $u$  are input
from the dynamical evolution of the turbulence and the total time
derivative $d_t$ includes advection with the compressible $E\times B$
velocity. 

The dynamical
equation for the impurity density  thus differs from the
dynamical equation for the density fluctuations Eq.~(\ref{eq:eqne}).  Typical
simplifications originating in the distinction between a background and
fluctuations are not possible to introduce for the impurity species.  
Finite inertia effects of the impurity ions enter 
through the ion-polarisation drift and finally all convection terms
need to be preserved, while for example parallel convection is
neglected in the dynamical equation for the density fluctuations.
Note that the equation for the impurity species do
not reflect the properties of the equilibrium as given by
Eqs. (\ref{eq:equil1}) and  (\ref{eq:equil2}) as the impurities  are
not assumed to contribute significantly to the establishment of quasi-neutrality.

Simulations were performed on a grid with $128 \times 512
\times 32$ points and dimensions $64 \times 256 \times 2 \pi$ in ${x,y,s}$.
Standard parameters for the runs were $\widehat{\mu}=5$, $q = 3$,  magnetic
shear $\widehat{s}=1$ (appearing only in the geometrical setup of the
simulations \cite{Scott:1997:1} ), and
$\omega_{B}=0.05$, with the viscosities set to
$\mu_\omega=\mu_n=0.025$, corresponding to typical edge parameters of
large fusion devices.

In the saturated turbulent state the equilibrium density gradient is
weakly flattened. Strong deviations from the initially specified 
density gradient are, however,  prevented by a feed-back mechanism using
two damping layers at the inner and outer radial boundary. This forces
the flux-surface averaged density to stay close to its initially
specified equilibrium profile.

To investigate impurity diffusion in fully developed, quasi-stationary
turbulence 
we  numerically solve  Eqs.~(1) and let initial perturbations  develop until a state of
saturated, quasistationary turbulence is reached.
The impurities are then released into  the
turbulence and convected around by the turbulence according to the
evolving turbulent velocity field. The initial impurity density $n_{imp}$  is chosen as a radially
localized Gaussian added to a constant impurity background density. 
For some runs the impurity density  was additionally
localized along the magnetic field lines, that is in coordinate $s$, 
to investigate the effect of
parallel convective transport.  Here we choose to
investigate the behavior of massless impurities.
 The coupled system of bulk plasma turbulence and impurities 
is evolved  until
significant mixing of the impurities  has been achieved and initial
transient effects have decayed. For each
parameter several runs
are performed to increase the statistical significance of the results.

From the transport of passive fields in compressible fluid
turbulence it is well known that the passive quantity reveals a much flatter
fluctuation spectrum than the turbulent energy spectrum, moreover  the
passive scalar tends to accumulate in the contracting regions of the
turbulent flow field \cite{Falkovich:Gawdzki:Vergassola:2001}. Thus,
the impurity density after some time not only exhibits strong
gradients, but locally the initial value of the impurity density can
be exceeded. To be able to handle these effects  within the given
limited resolution the diffusive term in
Eq.~(\ref{eq:impurity}) was chosen as $\mu_{imp} = 5 \mu_n$

\begin{figure}[H]
\centering \includegraphics[width=.85\textwidth]{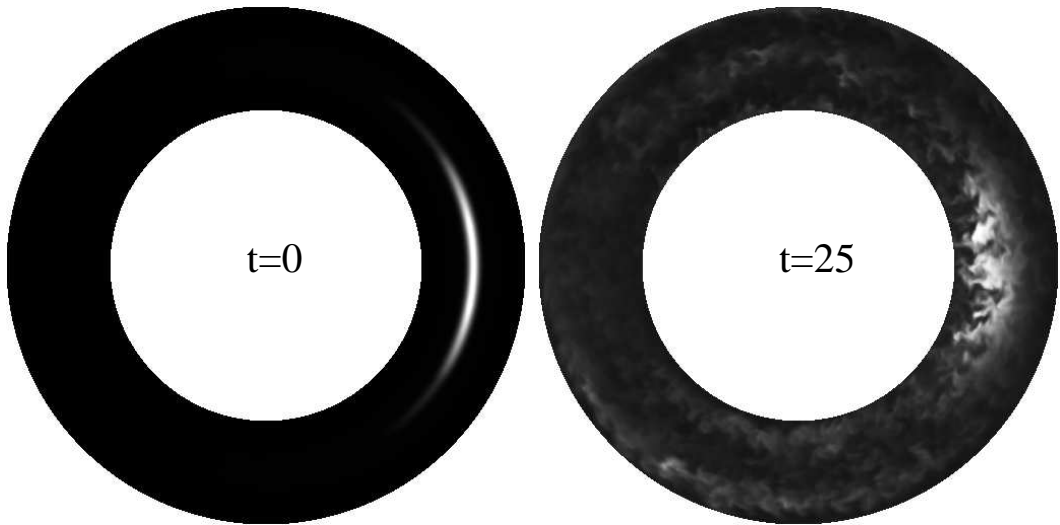}
\caption{Impurity distribution projected onto a poloidal cross-section
  (note: radial dimension not to scale), white: high impurity density
  and dark: low impurity density. Left: initial distribution,
  localized along the magnetic field on the outboard midplane (at $s= 0$ . Right: after 25 time
  units corresponding to about $100 \mu \mbox{s}$. Parallel transport is slow compared to radial
  transport. The inward pinch effect is clearly visible. 
}
\label{fig:local}
\end{figure}

A prominent feature of the impurity behavior is 
the weak parallel convective transport  compared to the radial turbulent
transport. The reason is that the impurities are convected  in the
parallel direction by the fluctuating parallel ion
speed $u$ which is small $u  \approx 0.01 $ compared to a radial
velocity that is of order one.  This is clearly  observed in
Fig. \ref{fig:local}, which shows the impurity density
projected onto a geometrically poloidal cut. The projection roughly
translates the variation in $s$ to a poloidal variation. $s=0$
corresponds to the outboard midplane, $s= \pi/2$ to the upper, $s = -
\pi/2$ to the lower side and finally  $s= \pm \pi$ corresponds to
the high field side. 
The radial
extend of the simulation domain has been stretched to allow better visualization.
The initial impurity density is
localized at the outboard midplane, corresponding to a parallel
localization in the flux tube geometry. No significant parallel flow of
the impurity density is observed, while significant radial mixing
occurs. Parallel compressional effects are
however visible and arrange for finite passive density gradients at
the high field side. Moreover an inward pinch effect is clearly observed at
the outboard midplane. 

\begin{figure}[H]
\centering \includegraphics[width=.85\textwidth]{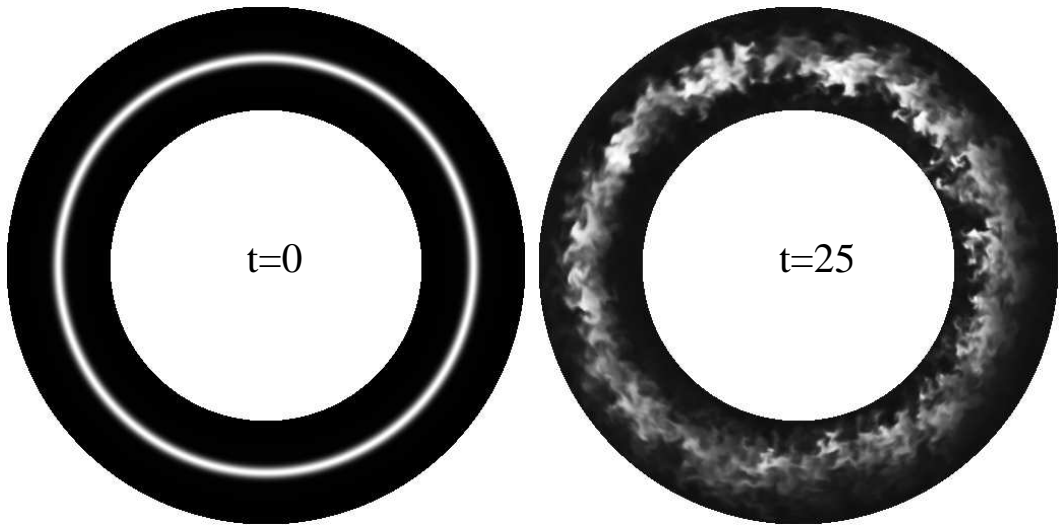}
\caption{Impurity distribution projected onto a poloidal cross-section
  (note: radial dimension not to scale). Left: initial distribution. Right: after 25 time
  units. The inward pinch effect on the outboard midplane and outward
  convective transport on the high field side (inboard midplane)  is obvious. 
}
\label{fig:ring}
\end{figure}

By starting from an initial impurity distribution
that is  homogeneous along $s$, this pinching  velocity is seen to
lead to a shift towards the torus axis of the impurity density compared to the
initial condition (Figure \ref{fig:ring}).

\begin{figure}[H]
\centering  \includegraphics[width=.85\textwidth]{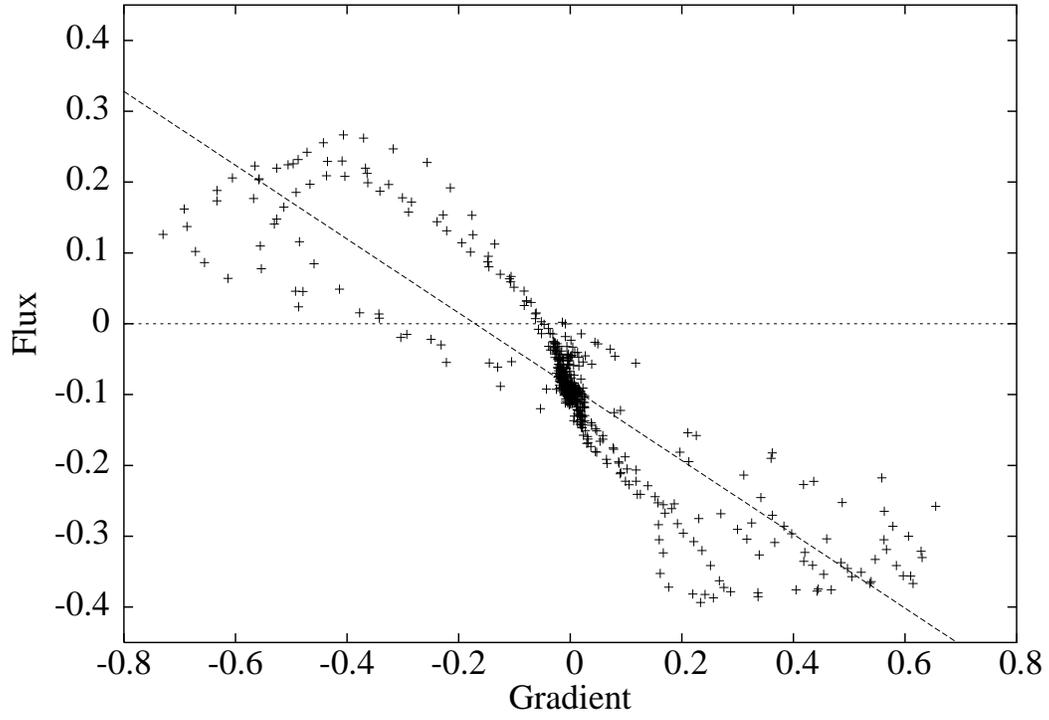}
\caption{Typical scatter plot (at high field side) of the flux versus gradient  with linear fit.}
\label{fig:visc_exb}
\end{figure}

\begin{figure}[H]
 \centering \includegraphics[width=.85\textwidth]{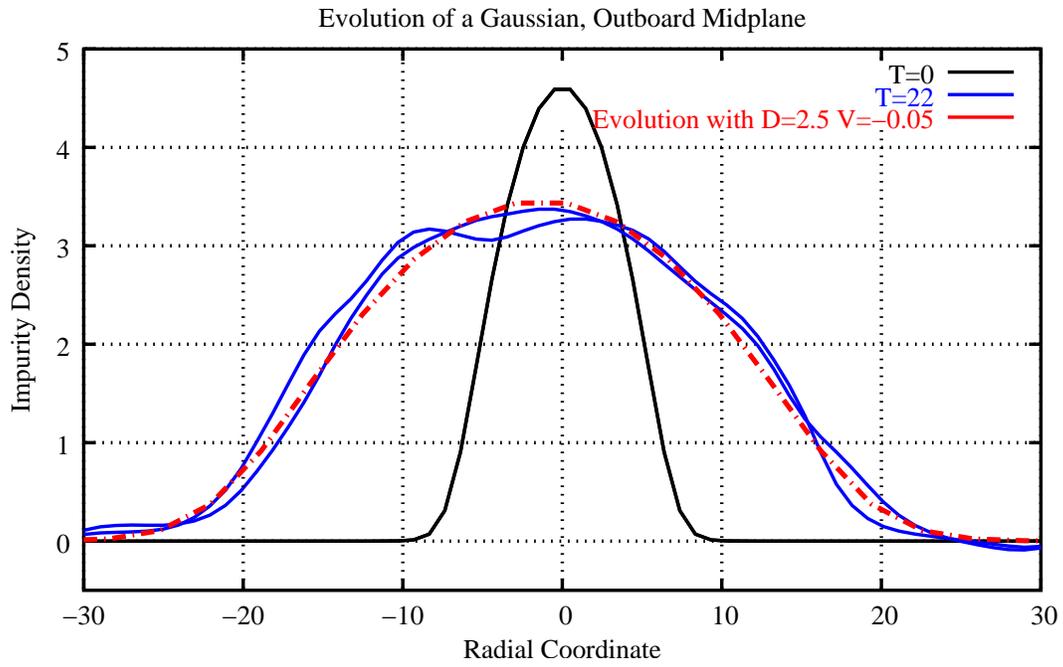}
\caption{Impurity density averaged along $y$ at the outboard midplane
  and compared to the expected evolution of a Gaussian from the fitted
  coefficients $D$ and $V$. }
\label{fig:curv_pinch}
\end{figure}

For a more quantitative description of this behavior, the
flux $\Gamma$ of the impurity ion species 
can in lowest order be expressed by a diffusion coefficient $D$ and a
convective velocity $V$, which is associated to a pinch effect:
$$
\Gamma_{y} (s) = -D(s) \partial_{x} <n>_{y} + V(s) <n>_{y}\,.
$$
The turbulence is radially homogeneous and so
there is no radial dependence of $D$ and $V$. Averages are taken along
the periodic  $y$ direction. We obtain these values for
each value of $s$ along the magnetic field lines and thus at different
poloidal positions. 
 From
a scatter plot of  $ \Gamma(r)/<n>_{y}  $ versus $\partial_x <\ln n>_{y}$ the
parameters $D(s)$ and $V(s)$ can be obtained. These are standard parameters
used in  modeling and in evaluation of transport experiments.  The fitting procedure is carried out using a
standard  nonlinear least-squares  Marquardt-Levenberg algorithm
\cite{Marquardt:1963} as implemented in the GNUPLOT software.
Figure \ref{fig:visc_exb} shows such a typical scatter plot with a
fitted linear relation between the two quantities indicating that
while there are significant deviations from a linear relationship,
the decomposition of the transport into $D$ and $V$
has some merit. The comparison of the 
evolution of the impurity density profile with the analytical
evolution of the profile using the values for $D$ and $V$, obtained
from an analysis as shown in Figure \ref{fig:visc_exb}
is depicted in Figure \ref{fig:curv_pinch}.
 
\begin{figure}[H]
 \centering \includegraphics[width=.85\textwidth]{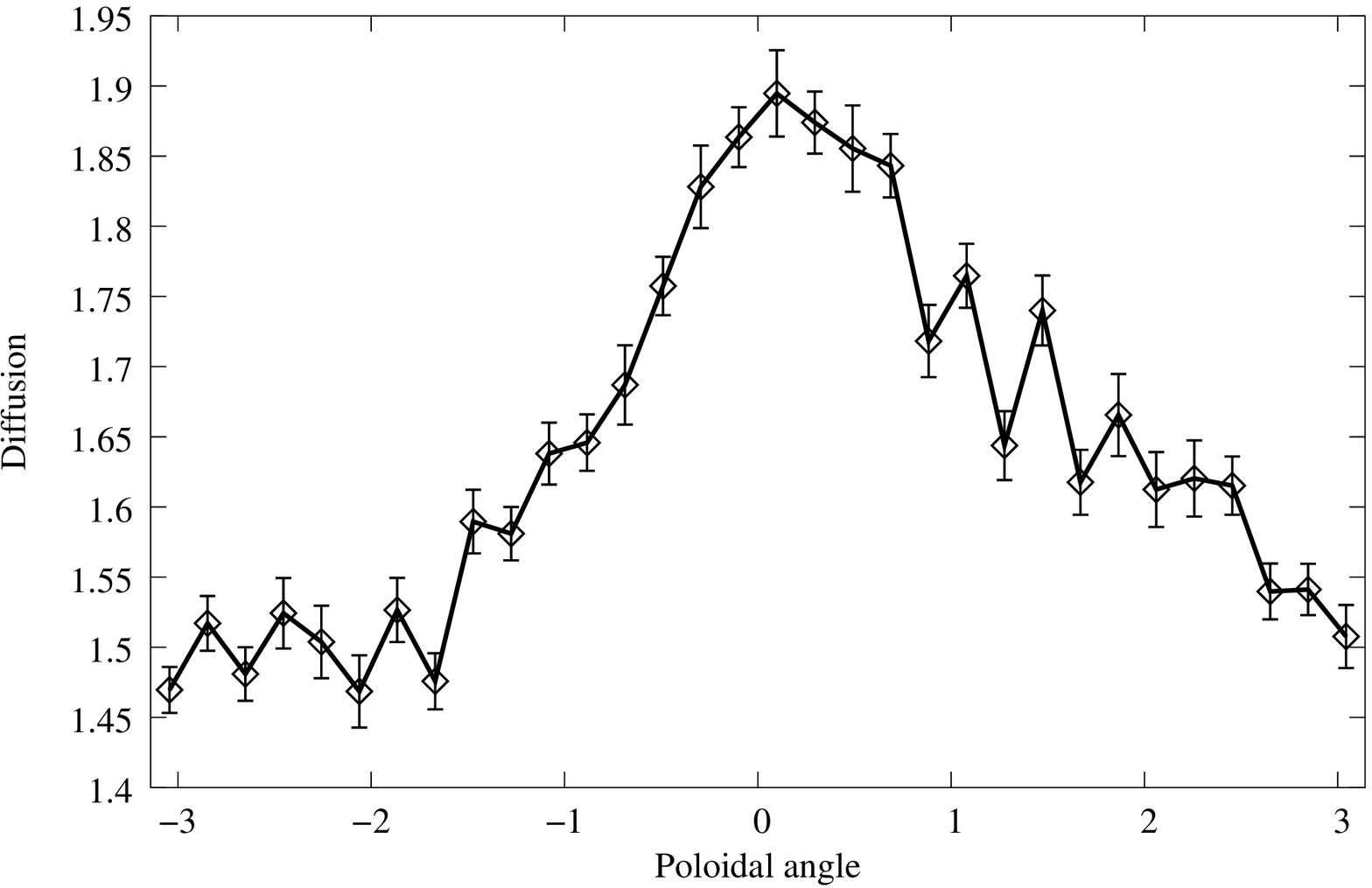}
\centering \includegraphics[width=.85\textwidth]{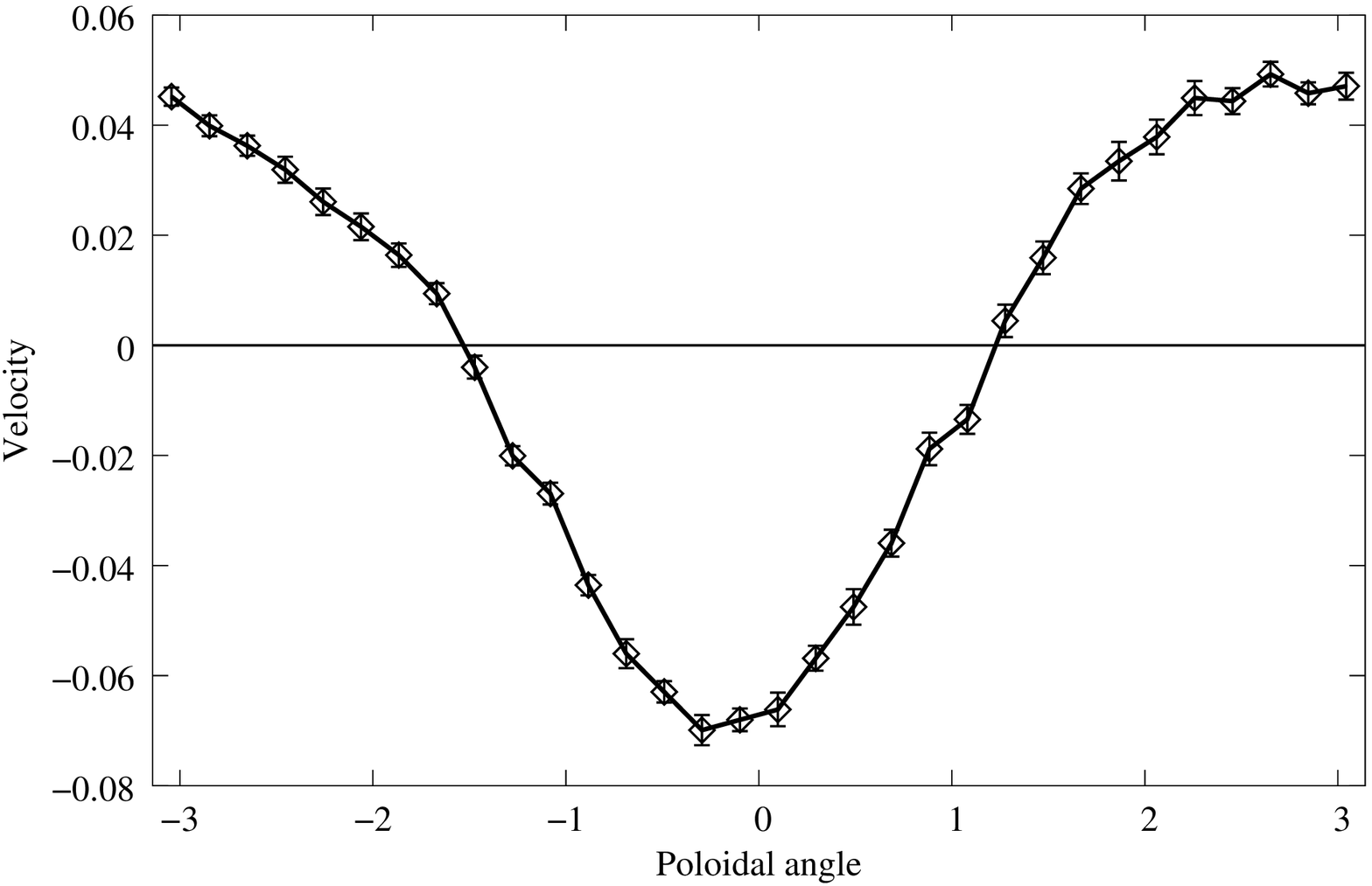}
\caption{Impurity diffusion  $D$ (top)  and pinch velocity  $V$ over
  poloidal position angle with indicated error-bars. 
The observed asymmetry is due to turbulence
  spreading in the diamagnetic direction.  }
\label{fig:vd}
\end{figure}

The poloidal dependence of 
diffusion and effective convection is rather strong and depicted in
Fig.~\ref{fig:vd}.
The effective convective  velocity $V(s)$ changes sign and is at the high
field side directed outwards.  This pinching velocity is due to normal
curvature and can be consistently explained in the framework of Turbulent
EquiPartition (TEP)
\cite{Nycander:Yankov:1995,Naulin:Rasmussen:Nycander:1998}. 
In the absence  of parallel convection, finite
mass effects and diffusion  Eq.~(\ref{eq:impurity})  has the following
approximate Lagrangian invariant
\begin{equation}
L(s) = \ln n_{imp} + \omega_B cos(s) x - \omega_B sin(s) y\;.
 \end{equation}

Its spatial  homogenization on each drift plane $<L(s)>_y = const(s)$ by the
turbulence indicates that at the
outboard midplane ($s=0$) the impurites are effectively convected
radially inward leading to a radially gradient ($ < \ln n_{imp}>_y
\propto  const.   -\omega_B  x   $) while at the high field side they are
effectively convected radially outward ($ < \ln n_{imp}>_y
\propto   const. +  \omega_B  x   $).
One should note that this effective inward or outward convection
is not found as an average $E \times B$ velocity, but is mitigated
by the effect of  homogenization of $L$ under the action of the turbulence. 
The strength of
the ``pinch'' effect is consequently proportional to the mixing properties of the
turbulence and thus to lowest order has an additional variation that  scales
with the measured turbulent diffusion, so that $V(s) \propto -\cos (s)
D(s) $. The slight ballooning in the
turbulence level thus causes  the inward flow  on the outboard midplane to
be stronger
than the effective outflow on the high-field side. Averaged over a flux surface
and assuming poloidally constant impurity density  a net impurity
inflow results. The net pinch is directly proportional to the diffusion
coefficient $D$ in agreement with experimental observations \cite{Perry:Brooks:Content:etal:1991}. 
Translated to  dimensional values for typical large tokamak edge parameters we
obtain $D(s) \propto 1.5 - 2.0 \; {\mbox{m}^2 }/{\mbox{s}}$ and $V(s)
\propto +60 - -80  \; {\mbox{m}}/{\mbox{s}}$ and an flux-surface averaged inward
convection velocity of $<V> = - 0.4 \; {\mbox{m}}/{\mbox{s}}$.  
Locally at the outboard midplane  values of $V(s=0)/D(s=0) \sim -40
\mbox{m}^{-1}$ are found, in rough agreement with experimental values
\cite{Dux:2003}.

Furthermore we  observe a slight shift of the peak diffusion
coefficient in the diamagnetic direction. This effect is likely due to the
spreading of the turbulence \cite{Lin:Hahm:2004} in that  direction leading to
enhanced turbulence levels slightly upward from the outboard midplane
and consequently to an up-down asymmetry.

The strong peaking of the impurity density, the poloidal dependence of
the transport coefficients and the slow parallel diffusion of the
impurities thus make it necessary to apply  at least two dimensional
modeling.  Both a  poloidally varying diffusion coefficient  and effective convection velocity 
should be used in transport codes to describe impurity density evolution. 
The observed impurity pinch in the edge plasma region can be explained
by turbulent equipartition without invoking arguments from
neoclassical transport theory. 

This work was supported by the Danish Center for Scientific Computing
through grants no.\ CPU-1101-08 and CPU-1002-17. 
Discussions with K.D. Zastrow and X. Garbet during a stay of the
author at JET are gratefully acknowledged. 


\newpage

\newpage

\newpage

\newpage

\newpage

\end{document}